\documentclass[runningheads]{llncs}
\usepackage{amssymb}
\usepackage{amsmath}
\usepackage[disable]{todonotes}
\usepackage{hyperref}
\usepackage[vlined,ruled,noend]{algorithm2e}
\usepackage{bussproofs}
\usepackage{mathtools}
\usepackage{tikz}
\usepackage{graphicx}
\usepackage{csquotes}
\usepackage{cite,doi}
\def\orcid#1{{\href{http://orcid.org/#1}{\protect\raisebox{-1.25pt}{\protect\includegraphics{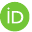}}}}}
\def\doi#1{\href{https://doi.org/\detokenize{#1}}{\url{https://doi.org/\detokenize{#1}}}}

\usepackage{xcolor}
\usepackage{pgfplots}
\usepackage{subcaption}

\newif\ifextendedVersion

\extendedVersiontrue %

\ifextendedVersion
\title{\Evonne: Interactive Proof Visualization for Description Logics (System Description) ---
Extended Version}
\else
\title{\Evonne: %
Interactive Proof Visualization for Description Logics %
(System Description)}
\fi

\titlerunning{Interactive Proof Visualization for DLs}

\author{
Christian~Alrabbaa\inst{1}
\orcid{0000-0002-2925-1765}
\and
Franz~Baader\inst{1}
\orcid{0000-0002-4049-221X}
\and
Stefan~Borgwardt\inst{1}
\orcid{0000-0003-0924-8478}
\and
Raimund~Dachselt\inst{2}
\orcid{0000-0002-2176-876X}
\and
Patrick~Koopmann\inst{1}
\orcid{0000-0001-5999-2583}
\and
Juli\'an~M\'endez\inst{2}
\orcid{0000-0003-1029-7656}
}
\authorrunning{C.\ Alrabbaa, F.\ Baader, et al.}
\institute{Institute of Theoretical Computer Science, TU Dresden, Germany \and
Interactive Media Lab Dresden, TU Dresden, Germany
\email{firstname.lastname@tu-dresden.de},
\email{julian.mendez2@tu-dresden.de}}
\newcommand{\ttodo}[4]{\ifthenelse{\equal{#1}{inline}}{\todo[inline, author=#2, color =
#3]{#4}}{\todo[color=#3]{#2: #4}}}

\newcommand{\wrt}{w.r.t.\ }

\newcommand{\ie}{i.e.\ }
\newcommand{\eg}{e.g.\ }
\newcommand{\cf}{cf.\ }

\newlength{\myl}
\newcommand{\longsquigarrow}[1]{
    \settowidth{\myl}{$~_{#1}$}
    \raisebox{-0.01cm}{\xymatrix@C=\myl{
            {}\ar@{~>}[r]^{~_{#1}}&{}
        }
    }
}

\def\define#1#2#3%
{%
\renewcommand*{\do}[1]{%
\expandafter\newcommand\csname
#1\endcsname{#2}
}
\docsvlist{#3}
}

\define{#1}
{{\text{\upshape{\textsc{#1}}}}\xspace}
{NP,coNP,PSpace,ExpTime,ExpSpace,NLogTime,NExpTime,LogSpace}

\newcommand{\Jmc}{\ensuremath{\mathcal{J}}\xspace}

\newcommand{\Omc}{\ensuremath{\mathcal{O}}\xspace}
\newcommand{\Pmc}{\ensuremath{\mathcal{P}}\xspace}
\newcommand{\Umc}{\ensuremath{\mathcal{U}}\xspace}

\newcommand{\ALCOI}{\ensuremath{\mathcal{ALCOI}}\xspace}
\newcommand{\EL}{\ensuremath{\mathcal{E}\hspace{-0.1em}\mathcal{L}}\xspace}
\newcommand{\ALCH}{\ensuremath{\mathcal{ALCH}}\xspace}
\newcommand{\ELH}{\ensuremath{\mathcal{E}\hspace{-0.1em}\mathcal{LH}}\xspace}

\newcommand{\exFont}[1]{\ensuremath{\mathsf{#1}}\xspace}

\newcommand{\Lethe}{\textsc{Lethe}\xspace}
\newcommand{\Elk}{\text{\upshape{\textsc{Elk}}}\xspace}

\newcommand{\Fame}{\textsc{Fame}\xspace}

\newcommand{\Evonne}{\textsc{Evonne}\xspace}

\newcommand{\evUp}{\includegraphics[width=.8em]{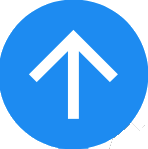}\xspace}
\newcommand{\evDown}{\includegraphics[width=.8em]{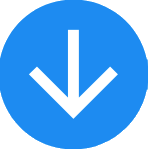}\xspace}
\newcommand{\evLeft}{\includegraphics[width=.8em]{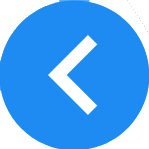}\xspace}
\newcommand{\evExplain}{\includegraphics[width=.8em]{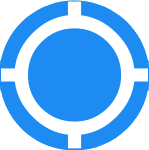}\xspace}

\newcommand{\Heur}{\texttt{HEUR}\xspace}
\newcommand{\SymMin}{\texttt{SYMB}\xspace}
\newcommand{\SizeMin}{\texttt{SIZE}\xspace}

\newcommand{\GPFP}{\texttt{GPFP}\xspace}
\newcommand{\GPS}{\texttt{GPS}\xspace}
\newcommand{\IPS}{\texttt{IPS}\xspace}
\newcommand{\Def}{\texttt{DEF}\xspace}

\newcommand{\AS}{\texttt{AS}\xspace}
\newcommand{\PS}{\texttt{PS}\xspace}
\newcommand{\STATE}{\texttt{S}\xspace}
\newcommand{\SVT}{\texttt{SVT}\xspace}

\tikzset{marks/.style={only marks, mark size = 1pt, solid, opacity=.7}}

\newcommand{\diagramFontSize}{\small}

\newcommand{\plotFBlog}[6]{
    \diagramFontSize
    \begin{tikzpicture}
        \begin{axis}[height=0.43\textwidth,
            width=0.43\textwidth,
            title={#1},
            xlabel={#2},
            ylabel={#3},
            xmode=log,
            ymode=log,
            every axis title/.append style={
                at={(0.5,.95)}
            },
            every axis y label/.style={
                at={(ticklabel cs:0.5)},rotate=90, anchor=near ticklabel,
            },
            every axis x label/.style={
                at={(ticklabel cs:0.5)}, anchor=near ticklabel,
            },
            xmin=99,
            xmax=#4,
            ymin=99,
            ymax=#4,
            grid=major,
            domain = 100:#4,
            xtick scale label code/.code={},
            ytick scale label code/.code={},
            xtick distance=10,
            ytick distance=10,
            ]
            \addplot[draw=black,samples at={100,1000000}] {x};
            \addplot[
                marks,
                green!60!black!60!,
                scatter,
                visualization depends on={value \thisrowno{2} \as \count},
                scatter/@pre marker code/.code={
                \def\markopts{
                    mark size=1+\count/10
                }
                \expandafter\scope\expandafter[\markopts]
                },
                scatter/@post marker code/.code={
                    \endscope
                }
            ] table [x index = 0, y index = 1, col sep = comma] {#5};
            \addplot[
                marks,
                mark = triangle*,
                red!60!black!60!,
                scatter,
                visualization depends on={value \thisrowno{2} \as \count},
                scatter/@pre marker code/.code={
                \def\markopts{
                    mark size=1+\count/10
                }
                \expandafter\scope\expandafter[\markopts]
                },
                scatter/@post marker code/.code={
                    \endscope
                }
            ] table [x index = 0, y index = 1, col sep = comma]  {#6};
        \end{axis}
    \end{tikzpicture}
}

\newcommand{\plotFB}[9]{
    \diagramFontSize
    \begin{tikzpicture}
        \begin{axis}[height=0.43\textwidth,
            width=0.43\textwidth,
            title={#1},
            xlabel={#2},
            ylabel={#3},
            every axis title/.append style={
                at={(0.5,.95)}
            },
            every axis y label/.style={
                at={(ticklabel cs:0.5)},rotate=90, anchor=near ticklabel,
            },
            every axis x label/.style={
                at={(ticklabel cs:0.5)}, anchor=near ticklabel,
            },
            xmin=0,
            xmax=#4,
            ymin=0,
            ymax=#4,
            domain = 0:#4,
            grid=major,
            xtick scale label code/.code={},
            ytick scale label code/.code={},
            xtick distance=10,
            ytick distance=10,
            legend style={font=\smaller},
            legend pos=#9,
            legend cell align={left},
            ]
            \addplot[draw=black] {x};
            \addplot[
                marks,
                green!60!black!60!,
                scatter,
                visualization depends on={value \thisrowno{2} \as \count},
                scatter/@pre marker code/.code={
                \def\markopts{
                    mark size=1+\count/10
                }
                \expandafter\scope\expandafter[\markopts]
                },
                scatter/@post marker code/.code={
                    \endscope
                }
            ] table [x index = 0, y index = 1, col sep = comma] {#5};
            \addplot[
                marks,
                mark = triangle*,
                red!60!black!60!,
                scatter,
                visualization depends on={value \thisrowno{2} \as \count},
                scatter/@pre marker code/.code={
                \def\markopts{
                    mark size=1+\count/10
                }
                \expandafter\scope\expandafter[\markopts]
                },
                scatter/@post marker code/.code={
                    \endscope
                }
            ] table [x index = 0, y index = 1, col sep = comma]  {#6};
            \legend{,#7,#8}
        \end{axis}
    \end{tikzpicture}
}

\newcommand{\plotR}[7]{
    \diagramFontSize
    \begin{tikzpicture}
        \begin{axis}[height=0.43\textwidth, 
            width=0.43\textwidth,
            title={#1},
            xlabel={#2},
            ylabel={#3},
            every axis title/.append style={
                at={(0.5,.95)}
            },
            every axis y label/.style={
                at={(ticklabel cs:0.5)},rotate=90, anchor=near ticklabel,
            },
            every axis x label/.style={
                at={(ticklabel cs:0.5)}, anchor=near ticklabel,
            },
            xmin=0,
            xmax=100,
            ymin=0,
            ymax=100,
            domain = 0:100,
            grid=major,
            xtick scale label code/.code={},
            ytick scale label code/.code={},
            ]
            \addplot[marks, green!60!black!60!] table [x index = 0, y index = 1, col sep=comma]  
            {#4}; 
            \addplot[marks,mark = diamond*, yellow!60!black!60!] table [x index = 0, y index = 1, col 
            sep=comma]  {#5};
            \addplot[marks,mark = star, blue!60!black!60!] table [x index = 0, y index = 1, col 
            sep=comma]  
            {#6};
            \addplot[marks, mark = triangle*, red!60!black!60!] table [x index = 0, y index = 1, col 
            sep=comma]  {#7};
        \end{axis}
    \end{tikzpicture}
}

\newcommand{\plotC}[9]{%
    \def\mTitle{#1}%
    \def\mXLabel{#2}%
    \def\mYLabel{#3}%
    \def\mXMax{#4}%
    \def\mFilea{#5}%
    \def\mFileb{#6}%
    \def\mFilec{#7}%
    \def\mFiled{#8}%
    \def\mLegPos{#9}%
    \plotCArgs%
}

\newcommand{\plotCArgs}[4]{%
    \def\mLegVala{#1}%
    \def\mLegValb{#2}%
    \def\mLegValc{#3}%
    \def\mLegVald{#4}%
    \plotCActual%
}

\newcommand{\plotCActual}{
    \diagramFontSize
    \begin{tikzpicture}
        \begin{axis}[height=0.43\textwidth, 
            width=0.43\textwidth,
            title={\mTitle},
            xlabel={\mXLabel},
            ylabel={\mYLabel},
            every axis title/.append style={
                at={(0.5,.95)}
            },
            every axis y label/.style={
                at={(ticklabel cs:0.5)},rotate=90, anchor=near ticklabel,
            },
            every axis x label/.style={
                at={(ticklabel cs:0.5)}, anchor=near ticklabel,
            },
            xmin=0,
            xmax=\mXMax,
            ymin=0,
            ymax=\mXMax,
            domain = 0:\mXMax,
            grid=major,
            xtick scale label code/.code={},
            ytick scale label code/.code={},
            xtick distance=50,
            ytick distance=50,
            legend style={font=\smaller},
            legend pos=\mLegPos,
            legend cell align={left}
            ]
            \addplot[draw=black] {x};
            \addplot[marks, green!60!black!60!] table [x index = 0, y index = 1, col sep=comma]  
            {\mFilea}; 
            \addplot[marks,mark = diamond*, yellow!60!black!60!] table [x index = 0, y index = 1, col 
            sep=comma]  {\mFileb};
            \addplot[marks,mark = star, blue!60!black!60!] table [x index = 0, y index = 1, col 
            sep=comma]  
            {\mFilec};
            \addplot[marks, mark = triangle*, red!60!black!60!] table [x index = 0, y index = 1, col 
            sep=comma]  {\mFiled};
           \legend{,\mLegVala,\mLegValb,\mLegValc,\mLegVald}
       \end{axis}
    \end{tikzpicture}
}

\begin{document}

\maketitle

\begin{abstract}
Explanations for description logic (DL) entailments provide important support for the maintenance of large ontologies.
The ``justifications'' usually employed for this purpose in ontology editors pinpoint the parts of the ontology responsible for a given entailment.
Proofs for entailments make the intermediate reasoning steps explicit, and thus explain how a  consequence can actually be derived.
We present an interactive system for exploring description logic proofs, called
\Evonne, which visualizes proofs of
consequences for ontologies written in expressive DLs.
We describe
the methods used for computing those proofs, together with a feature called
\emph{signature-based proof condensation}. Moreover, we evaluate the quality of
generated proofs using real ontologies.
\end{abstract}

\section {Introduction}

Proofs generated by Automated Reasoning (AR) systems are sometimes presented to 
humans in textual form to convince them of the correctness of a theorem~\cite{DBLP:conf/cade/Horacek99,Fiedler05},
but more often employed as certificates that can automatically be checked~\cite{DBLP:conf/cade/Reger017}.
In contrast to the AR setting, where very long proofs may be needed to derive a deep mathematical 
theorem from very few axioms, DL-based ontologies are often very large, but proofs of a single 
consequence are
usually of a more manageable size.
For this reason, the standard method of explanation in description logic~\cite{DLtextbook} has long been to compute so-called \emph{justifications}, which point out a minimal set of source statements responsible for an entailment of interest. For example, the ontology editor Protégé\footnote{\url{https://protege.stanford.edu/}} supports the computation of justifications since 2008~\cite{DBLP:conf/semweb/HorridgePS08a}, which is very useful when working with large DL ontologies. Nevertheless, it is often not obvious why a given consequence actually follows from such a justification~\cite{DBLP:conf/semweb/HorridgePS10}. Recently, this explanation capability has been extended towards showing full \emph{proofs} with intermediate reasoning steps, but this is restricted to ontologies written in the lightweight DLs supported by the \Elk reasoner~\cite{ELK,ProtegeProofExplanation}, and the graphical presentation of proofs is very basic.

In this paper, we present \Evonne as an interactive system, for exploring DL proofs for description logic entailments, using the methods for computing small proofs presented in~\cite{LPAR20,CADE21}. Initial prototypes of \Evonne were presented in~\cite{DBLP:conf/dlog/AlrabbaaBDFK20,DBLP:conf/semweb/FlemischLAD20}, but since then, many improvements were implemented. While \Evonne does more than just visualizing proofs, this paper focuses on the proof component of \Evonne: specifically, we give a brief overview of the interface for exploring proofs, describe the proof generation methods implemented in the back-end, and present an experimental evaluation of these proofs generation methods in terms of proof size and run time. 
The improved back-end
uses Java 
libraries that extract proofs using various methods, such as from the \Elk calculus, or 
\emph{forgetting-based proofs}~\cite{LPAR20} using the forgetting tools \Lethe~\cite{LETHE} 
and \Fame~\cite{FAME} in a black-box fashion.
The new 
front-end is visually more appealing than the prototypes presented in~\cite{DBLP:conf/dlog/AlrabbaaBDFK20,DBLP:conf/semweb/FlemischLAD20}, and allows to inspect and explore proofs using various interaction techniques, 
such as zooming and panning, collapsing and expanding, text manipulation, and compactness 
adjustments.
Additional features include the minimization of the generated proofs according to various measures and the possibility to select a \emph{known signature} that is used to automatically hide parts of the proofs that are assumed to be obvious for users with certain previous knowledge. Our evaluation shows that proof sizes can be significantly reduced in this way, 
making the proofs more user-friendly.
Evonne can be tried online and downloaded at \url{https://imld.de/evonne}.
The version of \Evonne described in this paper, as well as the data and scripts used in our 
experiments, can be found at~\cite{alrabbaa_christian_2022_6560603}.

\section {Preliminaries}

We recall some relevant notions for DLs; for a detailed introduction, see~\cite{DLtextbook}.
DLs are decidable fragments of first-order logic (FOL) with a special, variable-free syntax,
and that use only unary and binary predicates, called \emph{concept names} and \emph{role names},
respectively.
These can be used to build %
complex \emph{concepts}, which correspond to first-order formulas with one free variable, and 
\emph{axioms} corresponding to first-order sentences.
Which kinds of concepts and axioms can be built depends on the expressivity of the used DL.
Here we mainly consider the light-weight DL~\ELH and the more expressive \ALCH.
We have the usual notion of FOL \emph{entailment} $\Omc\models\alpha$ of an axiom~$\alpha$ from a finite set of axioms \Omc, called an ontology. 
Of special interest are entailments of \emph{atomic CIs} (concept inclusions)
of the form~$A\sqsubseteq B$, where $A$ and $B$ are concept names.
Following~\cite{LPAR20}, we define \emph{proofs} of $\Omc\models\alpha$ as finite, acyclic, directed hypergraphs, where vertices~$v$ are labeled with axioms~$\ell(v)$ and hyperedges are of the form $(S,d)$, with $S$ a set of vertices and $d$ a vertex such that $\{\ell(v)\mid v\in S\}\models\ell(d)$; the leaves of a proof must be labeled by elements of~\Omc and the root by~$\alpha$.
In this paper, all proofs are \emph{trees}, \ie no vertex can appear in the first component of 
multiple hyperedges (see Fig.~\ref{fig:evonne-overview-magic}).

\section {The Graphical User Interface}

\begin{figure}[tb]
    \includegraphics[width=1\textwidth]{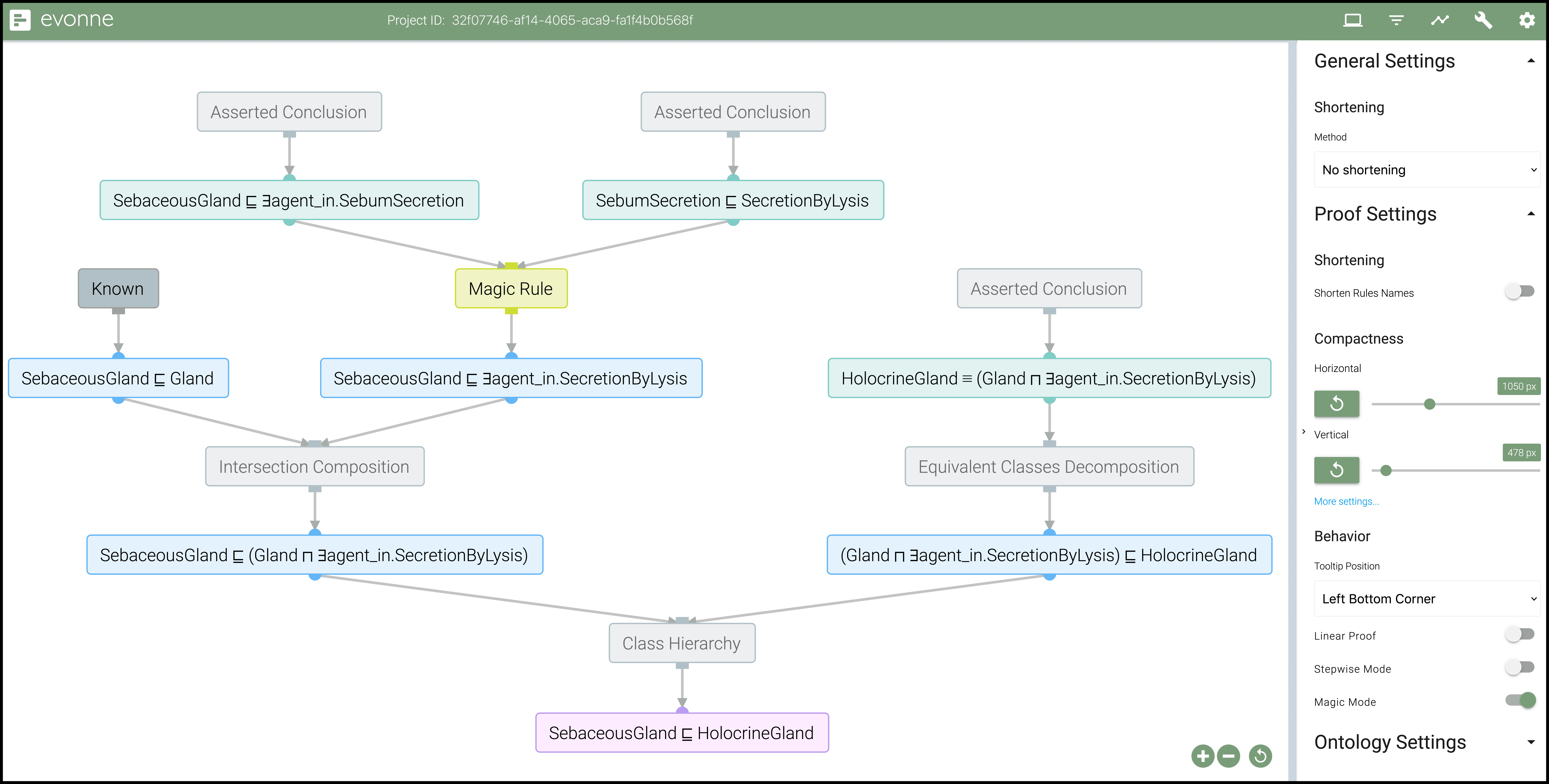}
    \caption{Overview of \Evonne~-~a condensed proof in the bidirectional layout}
    \label{fig:evonne-overview-magic}
\end{figure}

The user interface of \Evonne is implemented as a web application. To support users in understanding large proofs,
they are offered various layout options and interaction components.
The proof visualization is linked
to a second view showing the context of the proof in a relevant subset of the ontology. In this ontology view, interactions
between axioms are visualized, so that users can understand the context of axioms occurring in the proof. The user can also
examine possible ways to eliminate unwanted entailments in the ontology view. The focus of this system description, however,
is on the proof component: we describe how the proofs are generated and how users can interact with the proof visualization.
For details on the ontology view, we refer the reader to the workshop paper~\cite{DBLP:conf/dlog/AlrabbaaBDFK20}, 
where we also describe how \Evonne supports ontology repair.

\paragraph{Initialization.}
After starting \Evonne for the first time, users create a new project, for which they specify an 
ontology file.
They can then select an entailed atomic CI %
to be explained. The user can choose between different proof methods, and optionally
select a signature of \emph{known terms} (\cf Sec.~\ref{sec:proofGeneration}), which can be 
generated using the term selection tool Prot\'eg\'e-TS~\cite{TERM-SELECTION}.

\paragraph{Layout.}
Proofs are shown as graphs with two kinds of vertices: colored vertices for axioms, gray
ones for inference steps.
By default, proofs are shown using a \emph{tree layout}.
To take advantage of the width of the display when dealing with long axioms, it is possible to 
show proofs in a \emph{vertical layout}, placing axioms linearly below each other, with inferences represented through edges on the side (without the inference vertices).
It is possible to automatically re-order vertices to minimize
the distance between conclusion and premises in each step.
The third layout option is the \emph{bidirectional layout} (see 
Fig.~\ref{fig:evonne-overview-magic}), a tree 
layout where, initially, the entire proof is collapsed into a \emph{magic vertex} that links the conclusion 
directly to its justification, and from which individual inference steps can be pulled out and pushed 
back from both directions.

\paragraph{Exploration.}
In all views, each vertex is equipped with multiple functionalities for exploring a proof.
For proofs generated with \Elk, clicking on an inference vertex shows the inference rule used,
and the particular inference with relevant sub-elements highlighted in different colors.
Axiom vertices show different button (\evUp, \evDown, \evLeft, \evExplain) when hovered 
over. In the
standard tree layout, users can hide sub-proofs under an axiom
(\evDown). %
They can also reveal the previous inference step~(\evLeft) or the entire-sub-proof (\evUp).
In the vertical layout, the button (\evExplain) highlights and explains the inference of the current 
axiom.
In the bidirectional layout, the arrow buttons are used for pulling inference steps out of the magic 
vertex, as well as pushing them back in. 

\paragraph{Presentation.}
A \emph{minimap} allows users to keep track of the overall structure of the proof, thus
enriching the zooming and panning functionality.
Users can adjust width and height of proofs through the options side-bar.
Long axiom labels can be \emph{shortened} in two ways: either by setting a fixed size to all vertices, or 
by abbreviating names based on capital letters.
Afterwards, it is possible to restore the original labels individually.

\section {Proof Generation}\label{sec:proofGeneration}

To obtain the proofs that are shown to the user, we implemented different proof generation techniques, some of which were
initially described in~\cite{LPAR20}. For \ELH ontologies, proofs can be generated natively by the DL reasoner \Elk~\cite{ELK}. These proofs use rules from the calculus described in~\cite{ELK}. %
We apply the Dijkstra-like algorithm introduced in~\cite{DL20,CADE21} to compute %
a \emph{minimized proof} from the \Elk output.
This minimization can be done \wrt different measures, such as the size, depth, or weighted sum 
(where each axiom is weighted by its size), as long as they are \emph{monotone} and 
\emph{recursive}~\cite{CADE21}.
For ontologies outside of the \ELH fragment, we use the forgetting-based approach originally 
described in~\cite{LPAR20}, for which we now implemented two alternative algorithms for 
computing more compact proofs (Sec.~\ref{sec:fba-description}).
Finally, independently of the proof generation method, one can specify a signature of known terms.
This signature contains terminology that the user is familiar with, so that entailments using only those terms do not need to be explained. The condensation of proofs w.r.t.\ signatures
is described in Sec.~\ref{sec:sig-description}.

\subsection{Forgetting-Based Proofs}
\label{sec:fba-description}

In a forgetting-based proof, proof steps represent inferences on concept or
role names using a \emph{forgetting} operation.
Given an ontology $\Omc$ and a predicate name~$x$, the result $\Omc^{-x}$ of forgetting $x$ in $\Omc$ %
does not contain
any occurrences of~$x$, while still capturing all entailments of $\Omc$ that do not use $x$~\cite{FORGETTING}.
In a forgetting-based proof, an inference takes as premises a set
$\Pmc$ of axioms and has as conclusion some axiom $\alpha\in\Pmc^{-x}$ (where a particular forgetting
operation is used to compute $\Pmc^{-x}$). Intuitively, $\alpha$ is obtained from $\Pmc$ by performing
inferences on $x$.
To compute a forgetting-based proof, we have to forget the names occuring in the ontology one
after the other, until only the names occurring in the statement to be proved are left.
For the forgetting operation, the user can select between two
implementations: \Lethe~\cite{LETHE} (using the method supporting $\ALCH$) and
\Fame~\cite{FAME} (using the method supporting \ALCOI).
Since the space of possible inference steps is exponentially large, it is not feasible to minimize
proofs after their computation, as we do for \EL entailments, which is why we rely on heuristics and search algorithms to generate small proofs.
Specifically, we implemented three methods for computing forgetting-based proofs: \Heur tries 
to find proofs fast,
\SymMin tries to minimize the number of predicates forgotten in a proof, with the aim of 
obtaining proofs of small depth, and
\SizeMin tries to optimize the size of the proof. The heuristic method \Heur is described 
in~\cite{LPAR20}, and its implementation has not been changed since then.
The search methods \SymMin and \SizeMin are new 
\ifextendedVersion
(details can be found in the appendix).
\else
(details can be found in the extended version~\cite{thisExtended}).
\fi

\subsection{Signature-Based Proof Condensation}
\label{sec:sig-description}

When inspecting a proof over a real-world ontology, different parts of the proof
will be more or less familiar to the user, depending on their knowledge about the
involved concepts or their experience with similar inference steps in the past.
For CIs between concepts for which a user has application knowledge, they may not need to
see a proof, and consequently, sub-proofs for such axioms can be automatically hidden.
We assume that the user's knowledge is given in the form of a \emph{known signature}~$\Sigma$
and that axioms that contain only symbols from~$\Sigma$ do not need to be explained.
The effect can be seen in Fig.~\ref{fig:evonne-overview-magic} through the 
``known''-inference on the left, where
$\Sigma$ contains
$\exFont{SebaceousGland}$ and $\exFont{Gland}$.
The known signature is taken into consideration when minimizing the proofs, so that proofs are selected
for which more of the known information can be used if convenient. This can be easily integrated
into the Dijsktra approach described in~\cite{LPAR20}, by initially assigning to each axiom covered by $\Sigma$ a proof with a single vertex.

\section {Evaluation}

For \Evonne to be usable in practice, it is vital that proofs are computed efficiently and that they are not too large. An experimental evaluation of minimized proofs for \EL and forgetting-based proofs
obtained with \Fame and \Lethe is provided in~\cite{LPAR20}. We here present an evaluation of additional aspects:
1) a comparison of the three methods for computing forgetting-based proofs, and 2) an evaluation on the impact of signature-based proof condensation.
All experiments were performed on Debian Linux (Intel Core i5-4590, 3.30\,GHz, 23\,GB Java heap size).

\subsection{Minimal Forgetting-Based Proofs}
\label{sec:fba-eval}
\begin{figure}[tb]
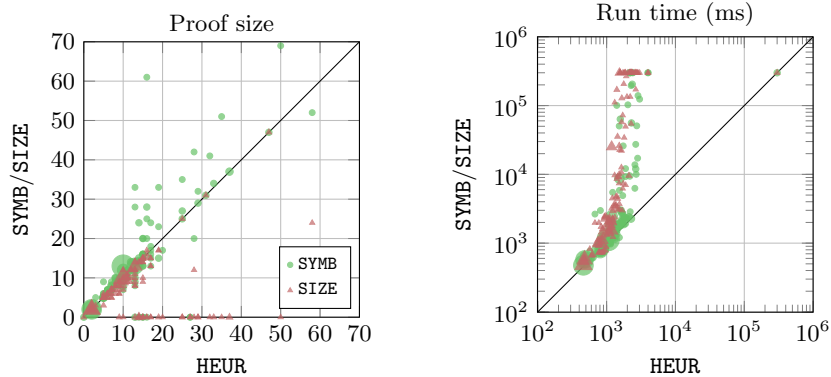

  \centering
  \plotFB{Proof size}{\Heur}{\SymMin/\SizeMin}{70}%
  {figures/fba/LETHE-orig-LETHE-symb-treesize-adj.csv}%
  {figures/fba/LETHE-orig-LETHE-size-treesize-adj.csv}%
  {\SymMin}{\SizeMin}{south east}
  \hfil
  \plotFBlog{Run time (ms)}{\Heur}{\SymMin/\SizeMin}{1000000}%
  {figures/fba/LETHE-orig-LETHE-symb-totaltime-adj.csv}%
  {figures/fba/LETHE-orig-LETHE-size-totaltime-adj.csv}%
  \caption{Run times and proof sizes for different forgetting-based proof methods. Marker size indicates how often each pattern occurred in the BioPortal snapshot. Instances that timed out were assigned size~$0$.}
  \label{fig:fba-eval}
\end{figure}

To evaluate forgetting-based proofs, we extracted $\ALCH$ ``proof tasks'' from the ontologies in the
2017 snapshot of
BioPortal~\cite{BIOPORTAL}. We restricted all ontologies to~\ALCH and collected all entailed 
atomic CIs~$\alpha$, for each of which we computed the union~$\Umc$ of all their
justifications. We identified pairs~$(\alpha,\Umc)$ that were isomorphic 
modulo
renaming of predicates, and kept only those patterns $(\alpha,\Umc)$ that contained at least one axiom not expressible 
in~\ELH. This was successful in 373 of the ontologies\footnote{The other ontologies could not be processed in this way within the memory limit.}
and resulted in 138 distinct
\emph{justification patterns} $(\alpha,\Umc)$, representing 327 different entailments in the BioPortal snapshot.
We then computed forgetting-based proofs for $\Umc\models\alpha$ with our three methods using \Lethe, with a 5-minute timeout.
This was successful for 325/327 entailments for the heuristic method (\Heur), 317 for the symbol-minimizing method (\SymMin),
and~279 for the size-minimizing method (\SizeMin). In Fig.~\ref{fig:fba-eval} we compare  the resulting \emph{proof sizes} (left) 
and the \emph{run times} (right), using \Heur as baseline (x-axis).
\Heur is indeed faster in most cases, but \SizeMin reduces proof size by~5\% on average compared to \Heur, which is not the case for \SymMin.
Regarding \emph{proof depth} (not shown in the figure), \SymMin did not outperform \Heur on average, while \SizeMin surprisingly yielded an average reduction of~$4\%$ compared to \Heur.
Despite this good
performance of \SizeMin for proof size and depth, for entailments that depend on many or complex axioms, computation times for both
\SymMin and \SizeMin become unacceptable, while proof generation with \Heur mostly stays in the
area of seconds.

\subsection{Signature-Based Proof Condensation}
\label{sec:sig-eval}

\begin{figure}[tb]
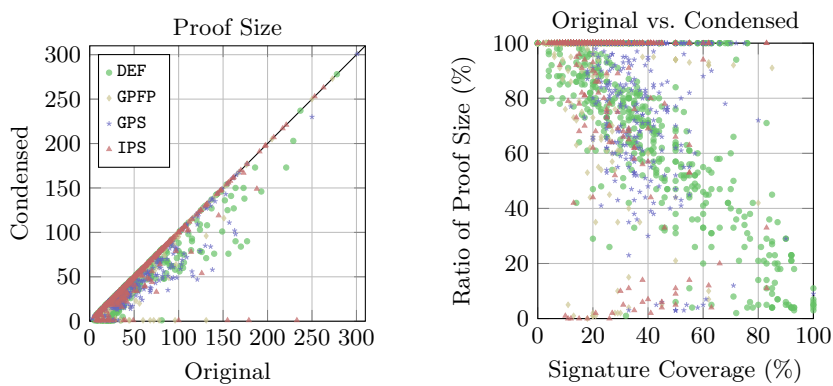

    \centering
    \plotC{Proof Size}{Original}{Condensed}{310}%
    {figures/treesize-comp.csv}%
    {figures/treesize-comp-2.csv}%
    {figures/treesize-comp-3.csv}%
    {figures/treesize-comp-4.csv}%
    {north west}{\Def}{\GPFP}{\GPS}{\IPS}
    \hfil
    \plotR{Original vs. Condensed}{Signature Coverage (\%)}{Ratio of Proof Size (\%)}%
    {figures/treesize-ratio-vs-coverage.csv}%
    {figures/treesize-ratio-vs-coverage-2.csv}%
    {figures/treesize-ratio-vs-coverage-3.csv}%
    {figures/treesize-ratio-vs-coverage-4.csv}
    \caption{Size of original and condensed proofs (left). Ratio of proof size depending on the 
    signature coverage (right).} %
    \label{fig:snomed-tree-size}
\end{figure}

To evaluate how much hiding  proof steps in a known signature decreases proof size in practice, we ran experiments on the large medical ontology SNOMED\,CT (International Edition, July 2020) that is mostly formulated in \ELH.\footnote{\url{https://www.snomed.org/}}
As signatures we used SNOMED\,CT \emph{Reference Sets},\footnote{\url{https://confluence.ihtsdotools.org/display/DOCRFSPG/2.3.+Reference+Set}} which are restricted vocabularies for specific use cases.
We extracted justifications similarly to the previous experiment, but did not rename predicates and considered only proof tasks that use at least~$5$ symbols from the signature, since otherwise no improvement can be expected by using the signatures.
For each signature, we randomly selected 500 out of 6.689.452 \emph{proof tasks} (if at least 500 existed).
This left the 4 reference sets \emph{General Practitioner/Family Practitioner} (\GPFP), \emph{Global Patient Set} (\GPS), \emph{International Patient Summary} (\IPS), and the one included in the SNOMED\,CT distribution (\Def).
For each of the resulting 2.000 proof tasks, we used \Elk~\cite{ELK} and our proof minimization approach to obtain (a)~a proof of minimal size and (b)~a proof of minimal size after hiding the selected signature.
The distribution of proof sizes can be seen in Fig.~\ref{fig:snomed-tree-size}.
In 770/2.000 cases, a smaller proof was generated when using the signature.
In 91 of these cases, the size was even be reduced to~1, \ie the target axiom used only the given signature and therefore nothing else needed to be shown.
In the other 679 cases with reduced size, the average \emph{ratio} of reduced size to original size was 0.68--0.93 (depending on the signature).
One can %
see that this ratio is correlated with the \emph{signature coverage} of the original 
proof (\ie the ratio of signature symbols to total symbols in the proof), with a weak or strong 
correlation depending on the signature ($r$ between~$-0.26$ and $-0.74$).
However, a substantial number of proofs with relatively high signature coverage could still not be reduced in size at all (see the top right of the right diagram).
In summary, we can see that signature-based condensation can be useful, but this depends on the proof task and the signature.
We also conducted experiments on the Galen 
ontology,\footnote{\url{https://bioportal.bioontology.org/ontologies/GALEN}} with comparable 
results 
\ifextendedVersion
(see the appendix).
\else
(see the extended version of this paper \cite{thisExtended}).
\fi

\section {Conclusion}

We have presented and compared the proof generation and presentation methods used in \Evonne, a visual tool for explaining entailments of DL ontologies.
While these methods produce smaller or less deep proofs, which are thus easier to present,
there is still room for improvements. Specifically, as the forgetting-based proofs do not provide the same degree of detail as the \Elk proofs, it would be desirable to also support methods for more expressive DLs that generate proofs with smaller inference steps. Moreover, 
our current evaluation focuses on proof size and depth---to understand how well \Evonne helps users to understand DL entailments, we would also need a qualitative evaluation of the tool with potential end-users. 
We are also working on explanations for non-entailments using counter\-models~\cite{DBLP:conf/ki/AlrabbaaHT21} and a plugin for the ontology editor Protégé that is compatible with the PULi library and Proof Explanation plugin presented in~\cite{ProtegeProofExplanation}, which will support all proof generation methods discussed here and more.\footnote{\url{https://github.com/de-tu-dresden-inf-lat/evee}}

\subsubsection{Acknowledgements}
This work was supported by the %
German Research Foundation (DFG)
in Germany’s Excellence Strategy: EXC-2068, 390729961 – Cluster 
of 
Excellence “Physics of Life” and EXC 2050/1, 390696704 – Cluster of Excellence “Centre for 
Tactile Internet” (CeTI) of TU Dresden, by DFG grant 389792660 as part of TRR 248 – CPEC,
and the DFG Research Training Group 
QuantLA, 
GRK 1763. %

\bibliographystyle{splncs04}

\newpage

\appendix

\ifextendedVersion
\section{Appendix}

\subsection{Algorithms Used for Computing Forgetting-Based Proofs}

We have implemented three algorithms for generating forgetting-based proofs, which
trade off computation time against the size of the proof.
The heuristic method is the recommended option, as it is the fastest and provides reasonably
good solutions (see Sec.~\ref{sec:fba-eval}). The others are more costly, and minimize
size and number of symbols eliminated, respectively. 
We recommend users to use the heuristic method first
to get an idea --- if the number of concept names occurring in the proof is not too large, 
they can then try out one of 
the alternative options to get a nicer proof.

\medskip

\textbf{Preliminaries.} Forgetting was already defined in the main text. The other technique we use
to compute minimal proofs is \emph{justifications}: given an ontology $\Omc$ and an axiom $\alpha$
s.t. $\Omc\models\alpha$, a justification for $\alpha$ in $\Omc$ is a subset-minimal $\Jmc\subseteq\Omc$
s.t. $\Jmc\models\alpha$~\cite{JUSTIFICATIONS}. In the worst case, there can be  exponentially many
justifications for an entailment. Our methods for computing forgetting-based proofs make use of the
black-box implementation for justifications provided by the OWL API~\cite{OWL-API}, using HermiT~\cite{HermiT} as a reasoner.
We use this functionality to extract one justification for a given entailment, and do not compute and
compare all possible justifications. This is a source of optimization we might investigate in 
future work.

Our implementations
support two libraries for forgetting in DLs, \Lethe~\cite{LETHE} and~\Fame~\cite{FAME}. \Lethe allows to set a relatively robust timeout
for computations---we use this feature and set the timeout to 2 seconds (both in the experiments and in the current prototype of \Evonne).
If \Lethe runs past this timeout, we consider the forgetting operation as failed.

\medskip

\textbf{Heuristic Method.}
In order to provide a proof for $\Omc\models A\sqsubseteq B$ ,
we step-wise forget all names in $\Omc$ except $A$ and $B$, ending up with an ontology that
contains either $A\sqsubseteq B$ itself or some axiom from which $A\sqsubseteq B$ easily
follows (\eg $A\sqsubseteq\bot$).
Forgetting is potentially an expensive operation, and the result of forgetting
a sequence of names may be triple-exponentially larger than the original
ontology~\cite{FORGETTING_FOUNDATIONS}. Even if this bound is rarely reached in practice~\cite{LETHE},
we need to keep the current set of axioms small in each step, which we do by computing justifications.
Another challenge is that forgetting a name may not always be successful,
or result in an ontology that cannot be expressed using standard DLs, in which case we have to
skip forgetting that name.
The heuristic forgetting-based proof generation method generates a sequence of ontologies
as follows \cite{LPAR20}:
\begin{enumerate}
 \item Set $i=0$, compute a justification $\Jmc_0$ of $\Omc$ for $A\sqsubseteq B$, and initialize $\Sigma$ to contain all concept and
 role names that occur in $\Omc$.
 \item Repeat the following steps while $\Jmc_i$ contains a name that is in $\Sigma\setminus\{A,B\}$:
 \begin{enumerate}
  \item Select such a name $x\in\Sigma$ following a heuristic and remove it from $\Sigma$.
  \item Attempt to compute $(\Jmc_i)^{-x}$.
  \item If successful, set $\Jmc_{i+1}$ to be a justification for $\alpha$ in $(\Jmc_{i})^{-x}$,
        otherwise, set $\Jmc_{i+1}=\Jmc_i$.
  \item Increment $i$.
 \end{enumerate}
\end{enumerate}

From the resulting sequence $\Jmc_1$, $\ldots$, $\Jmc_n$ of ontologies, we construct inference steps by using each axiom in $\Jmc_i\setminus\Jmc_{i-1}$ as a conclusion and a justification for it in $\Jmc_{i-1}$ as premises. If $A\sqsubseteq B\not\in\Jmc_n$, we furthermore add an inference with $A\sqsubseteq B$ as conclusion and $\Jmc_n$ as premises (recall that $\Jmc_n$ is already a justification of $A\sqsubseteq B$).

\medskip

The heuristic does not guarantee that we always find a proof that is optimal in size and/or
readability. To optimize size, we cannot simply apply a Dijsktra-like algorithm on the proofs generated by the heuristic
method, as we do for the proofs generated by \Elk. Without a heuristic, there are exponentially many orders in which the
names could be forgotten, in addition to the potentially exponential number of justifications we could select in each
step. Adding to this the cost of forgetting, this means that the set of all possible forgetting-based inference steps
cannot be precomputed in practice. Instead, in order to compute optimal proofs practically, we need to use
optimized search strategies that try to minimize the number of forgetting operations to be performed. We implemented two
such methods described in the following. In all methods, we assume a
\emph{deterministic selection of justifications}, that is, we do not consider
alternative justifications, though in reality there can be exponentially many.
This is to avoid further overhead, and to keep the focus of the optimization on
the forgetting operations. We also cache forgetting results to avoid recomputations. Since
the result of forgetting role names often hides many inferences, and would lead to proofs that are
small, but not easy to understand, we only forget concept names in the following methods. Instead,
the heuristic approach is used to complete the proof after all concept names have been eliminated
using the respective method.  However, syntactic simplifications performed by the forgetting
tools have the effect that role names often still get eliminated along the way.
For example, forgetting
$D$ from $\{C\sqsubseteq\exists r.D,C\sqsubseteq\forall r.\neg D\}$ would result in
$C\sqsubseteq \bot$ as the result of simplifying $C\sqsubseteq\exists r.\top\sqcap\exists r.\bot$.
\medskip

\textbf{Symbol-Minimizing Method. }
In the \emph{symbol-minimal forgetting-based proof generation} we minimize the number of names
that
are forgotten in total. Specifically, we explore different possible orders using a best-first
strategy: In each step, we precompute the next possible steps for $\Jmc_{i+1}$, and try them out
one after the other starting with the one that has the least number of remaining names. We keep
track of the
shortest successful sequence of names (and the corresponding proof) and cancel computation
branches
on which the number of forgotten names is larger.

\medskip

\textbf{(Weighted) Size-Optimizing Method.}
The \emph{size-optimized forgetting-based proof generation} optimizes for size of 
the proof rather than on the number of eliminated names. 
Specifically, it explores the space of possible inferences using the following recursive
algorithm $\texttt{Prove}(\Jmc,n)$
for creating a proof for $A\sqsubseteq B$ in $\Jmc$
of size at most $n$:
\begin{enumerate}
 \item If $n\le 0$, FAIL.
 \item If $A\sqsubseteq B\in\Jmc$, a proof with a single vertex is returned.
 \item If $A\sqsubseteq B\not\in\Jmc$, but all concept names in $\Jmc$ (except for $A$ and $B$) have
already been eliminated,
 return a proof consisting of a single inference step inferring $A\sqsubseteq B$
from a justification of $A\sqsubseteq B$ in $\Jmc$, if the justification has a size less than $n$.
 \item\label{step:init-m} Otherwise, : %
 \item\label{step:iterate} for every name $x\not\in\{A,B\}$ that can still be forgotten in $\Jmc$, attempt to compute $\Jmc^{-x}$,
 and process each successful $\Jmc^{-x}$ in order according to a heuristic:
 \begin{enumerate}
  \item Attempt to compute a proof for $\Jmc^{-x}\models A\sqsubseteq B$
with $\texttt{Prove}(\Jmc^{-x},n-m)$, where $m=\lvert \Jmc\setminus\Jmc^{-x}\rvert$ (the number of axioms that are not used in the proof anymore).
  \item If successful, extend the obtained proof using the axioms in $\Jmc$, and update the bound $n$ with the size of this proof in case it
  was smaller.
 \end{enumerate}
 \item If no proof generation was successful, FAIL.
 \item Otherwise, return the best proof constructed.
\end{enumerate}

The heuristic used to pick the next forgetting result in Step~\ref{step:iterate}
evaluates each forgetting result by its size times the number of names still occurring in
it---a rough approximation of the expected proof size.
To obtain the size-minimal forgetting-based proof for $\Omc\models\alpha$, we
call $\texttt{Prove}(\Jmc,n)$, where $n$ is some maximal value and $\Jmc$ a
justification for $\Omc\models\alpha$.

We also allow to use a different size measure, which
takes the sum of the sizes of the axioms occurring in the proof, resulting in the
\emph{weighted size-optimizing method.}

\subsection {Signature-Based Proof Condensation (Continued)}

\begin{figure}[tb]
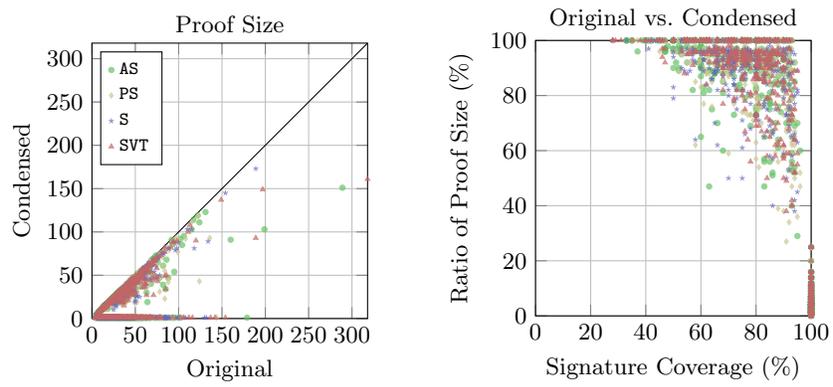

    \centering
    \plotC{Proof Size}{Original}{Condensed}{318}%
    {figures/figuresGalen/AbstractState-AbstractState-con-treesize.csv}%
    {figures/figuresGalen/ProcessState-ProcessState-con-treesize.csv}%
    {figures/figuresGalen/State-State-con-treesize.csv}%
    {figures/figuresGalen/SymbolicValueType-SymbolicValueType-con-treesize.csv}%
    {north west}{\AS}{\PS}{\STATE}{\SVT}
    \hfil
    \plotR{Original vs. Condensed}{Signature Coverage (\%)}{Ratio of Proof Size (\%)}%
    {figures/figuresGalen/AbstractState-AbstractState-con-treesize-ratio-vs-coverage.csv}%
    {figures/figuresGalen/ProcessState-ProcessState-con-treesize-ratio-vs-coverage.csv}%
    {figures/figuresGalen/State-State-con-treesize-ratio-vs-coverage.csv}%
    {figures/figuresGalen/SymbolicValueType-SymbolicValueType-con-treesize-ratio-vs-coverage.csv}
    \caption{Size of original and condensed proofs (left). Ratio of proof size depending on the 
        signature coverage (right).}
    \label{fig:galen-tree-size}
\end{figure}

We ran another experiment, similar to the one described in Sec.~\ref{sec:sig-eval}, with two 
main differences. First, we used another medical ontology called 
Galen.\footnote{\url{https://bioportal.bioontology.org/ontologies/GALEN}} Second, we generated the 
signatures based on the \emph{class hierarchy} of Galen.
In total, we computed $35$ signatures as follows: for every concept name in the top three levels of the 
class hierarchy, we extracted a module~\cite{DBLP:journals/jair/GrauHKS08}. Then we used the signature of each module to condense 
the proofs.
For proof generation, and analogous to the experiment of Sec.~\ref{sec:sig-eval}, we extracted 
proof tasks from Galen that use at least~$5$ symbols from a given signature. For each signature, we 
randomly selected at most $1000$ out of $452.070$ tasks. For $5$ signatures, less than $100$ tasks 
were extracted. Of the remaining signatures, $6$ had a range of tasks between $273$ and $693$. 
Each of the rest, with a total of $24$, had $1000$ tasks.
The condensation of proofs in this experiment exhibited a comparable behavior to the one illustrated in 
Fig.~\ref{fig:snomed-tree-size}. In Fig.~\ref{fig:galen-tree-size} we show the results for a sample of 
the signatures (derived from the modules for \emph{Abstract State} (\AS), \emph{Process State} (\PS), \emph{State} (\STATE), and 
\emph{Symbolic Value Type} (\SVT)).

\fi
\end{document}